\documentstyle[fleqn]{tp}

\input psfig

\def\d{{\rm d}}

\def\bfr{{\bf r}}
\def\vev#1{\left\langle#1\right\rangle}

\def\bfk{{\bf k}}

\def\bfv{{\bf v}}
\def\iras{{\it IRAS\/}}
\def\etal{{\it et al.\/}}
\def\kms{\ifmmode {\rm \ km \ s^{-1}}\else $\rm\, km \ s^{-1}$\fi}
\def\mpc{\ifmmode {\,h^{-1}\rm Mpc}\else$\,{h^{-1}\rm Mpc}$\fi}
\def\kpc{\ifmmode {\,h^{-1}\rm kpc}\else$\,{h^{-1}\rm kpc}$\fi}

\def\ltsima{$\; \buildrel < \over \sim \;$}
\def\simlt{\lower.5ex\hbox{\ltsima}}
\def\gtsima{$\; \buildrel > \over \sim \;$}
\def\simgt{\lower.5ex\hbox{\gtsima}}
\def\b{\item{$\bullet$\ }}

\begin{document}

\twocolumn[
\title{Large-Scale Flows as a Cosmological Probe}
\author{Michael A. Strauss$^{1,2}$ and Michael Blanton$^1$\\
{\it $^1$Princeton University Observatory, Princeton, NJ 08544 USA}\\
{\it $^2$Cottrell Scholar of Research Corporation}}
\vspace*{16pt}

ABSTRACT.\ We review the use of peculiar velocities of galaxies as a
probe of cosmological models.  We put particular emphasis on
comparison of the peculiar velocity and density fields, focussing on
the discrepancies between various recent analyses.  We 
discuss the limitations of the commonly used linear bias model, which
may lie at the heart of some of the current controversies in the
field. 
\endabstract]

\markboth{Michael A. Strauss and Michael Blanton}{Large-Scale Flows}

\small

\section{Introduction}
We believe for a large number of reasons (including those to be
outlined in this review) that we live in a universe which is dominated
by dark matter.  We see galaxies by the light of the stars that shine
within them; we can similarly infer the presence of large quantities
of hot gas through its X-ray emission, and cold gas through molecular
and atomic emission in the millimeter and radio portions of the
spectrum.  But the baryons that give rise to these emissions comprise
at most 10\% or so of the mass density of the universe. 

The study of large-scale structure asks 
how this dark matter is
distributed in space.  We cannot observe it directly,
but we can infer many of its properties due to its gravitational
influence on the matter around it.  We observe that the distribution
of galaxies is clustered, and make the assumption (which is at least
qualitatively justified, as we will see below) that the dark matter
distribution is clustered in a similar way.  A local overdensity in
mass will gravitationally attract the adjacent matter, 
disturbing the pure Hubble flow; similarly, a local underdensity
will gravitationally repel matter.  These additional motions, or {\em
peculiar velocities}, are observable.  Therefore, we can use peculiar
velocities to get an observational handle on the clustered component
of the dark matter.

  We make these notions concrete as follows. 

  Pure Hubble flow states that redshift is proportional to distance, 
\begin{equation} 
cz = H_0 r, 
\label{eq:Hubble} 
\end{equation}
and indeed, if we measure distances in units of \kms, and never make
reference to standard yardsticks calibrated in physical units (e.g.,
cm or Mpc), the value of $H_0$ is identically unity.  Peculiar
velocities are defined relative to the comoving expanding frame
implied by equation~(\ref{eq:Hubble}):
\begin{equation} 
cz = r + \hat \bfr \cdot \left[\bfv(\bfr) - \bfv({\bf 0})\right],
\label{eq:Hubble2} 
\end{equation}
where \bfv(\bfr) is the peculiar velocity field, and redshifts are
measured relative to the barycenter of the Local Group (as opposed to,
e.g., 
the CMB rest frame). 
 
In linear gravitational instability theory 
one can relate
the density fluctuations to the peculiar
velocities: 
\begin{equation} 
\bfv(\bfr) = {\Omega^{0.6} \over 4\,\pi} \int \d^3 \bfr'\,{
\delta(\bfr')\, (\bfr' - \bfr) \over |\bfr' - \bfr|^3},
\label{eq:v-integral} 
\end{equation}
where $\Omega$ is the Cosmological Density Parameter, and the density
fluctuation field is defined as: 
\begin{equation} 
\delta(\bfr) = {{\rho(\bfr) - \vev{\rho}} \over {\vev{\rho}}}.
\label{eq:delta} 
\end{equation}
 The integral on the right-hand side is just
proportional to the gravity vector.  As an aside, note that it is
manifestly not true that peculiar velocity is proportional to gravity
in the highly nonlinear regime: the gravity and velocity vectors of
the Earth's orbit are at right angles to one another!

We will also find it useful to express equation~(\ref{eq:v-integral}) in
differential form; by taking the divergence of both sides of the
equation, one finds: 
\begin{equation} 
\nabla \cdot \bfv(\bfr) = -\Omega^{0.6} \delta(\bfr).
\label{eq:del-dot-v}
\end{equation}

Note that peculiar
velocities, if measured 
in the linear regime, are a measure of the response
of galaxies to the gravitational influence of {\em all} matter, and
especially the dominant dark matter.  Therefore, we can measure the
large-scale distribution of dark matter directly from observations of
the peculiar velocity field. 

  Moreover, if we have an independent measurement of the large-scale
distribution  of {\em galaxies} from redshift surveys, and a model
that relates the galaxy and dark matter distribution, we can use
equation~(\ref{eq:v-integral}) or (\ref{eq:del-dot-v}) to
\begin{itemize} 
\b Test the gravitational instability paradigm from which these
equations were derived; 
\b Test the assumed model relating the distribution of the dark and
luminous matter; 
\b Measure the quantity $\Omega$ directly. 
\end{itemize}

\section{The Measurement of Peculiar Velocities}

Peculiar velocities manifest themselves through their modification to
the Hubble Law, equation~(\ref{eq:Hubble2}).  Therefore, with
independent measurements of redshifts and distances of a galaxy, we can measure
the {\em radial component} of its peculiar velocities.   Redshifts are
relatively easy to measure; a high-quality galaxy spectrum will show sharp
stellar absorption lines, and often strong emission lines that arise
from photo-ionized gas in H$\scriptstyle \rm II$ regions or an active
nucleus.  Distances are quite a bit more difficult; one needs to
identify a standard candle (i.e., a distance-independent way to
determine the absolute magnitude of the galaxy, or some well-defined
component of it), or a standard yardstick (i.e., a characteristic
physical size of the galaxy).  Comparison with the apparent magnitude
or angular size then yields a distance. 

Reviews of distance indicators used in peculiar
velocity work can be found in Jacoby \etal\ (1992), Strauss \& Willick
(1995, hereafter SW), and Willick (1998).  The Tully-Fisher (1977; TF)
relation  states that the luminosity and (inclination-corrected)
rotation velocity of a
spiral galaxy are related by a power law; expressing this in terms of
absolute magnitudes, we find: 
\begin{equation} 
M = A \log \Delta V + B.
\label{eq:TF} 
\end{equation}

Thus, observations of apparent magnitudes $m$ and rotation velocities
$\Delta V$ (via the H$\alpha$ line at 6563\AA\ with a long-slit
spectrograph placed along the major axis of the galaxy, or the width
of the 21 cm line in the radio), yield a distance.  The TF
relation is not perfect, and shows some appreciable scatter; distances
are measured typically to an accuracy of 15-20\%.  Moreover, the
constants $A$ and $B$ depend on the details of sample selection, the
band in which the galaxy photometry is done, and so on, and these
quantities need to be calibrated carefully for any given TF sample.
Since we measure the peculiar 
velocity as the difference of two rather large quantities
(equation~\ref{eq:Hubble2}), the signal-to-noise ratio {\em per
galaxy} of the peculiar velocity field is typically low, and is a
decreasing function of distance.  This means that a great deal of care
needs to be taken in doing TF work; when data are noisy, there are a
number of pernicious systematic effects that can creep into the
derived velocity field if one is not careful (cf., the discussion in
SW; Teerikorpi 1997). 

There are a number of other distance indicators that are used in
peculiar velocity work: 
\begin{itemize} 
\b Elliptical galaxies are observed to fall along the so-called
{\em fundamental plane} (Djorgovski \& Davis 1987; Dressler \etal\
1987), which relates the luminosity, the surface brightness, and the
velocity dispersion or color of galaxies.  This yields distances to an
accuracy of roughly 20\% in distance, comparable to the TF relation. 

\b Distances can also be measured to ellipticals using the method
of {\em surface brightness fluctuations} (Tonry \& Schneider 1988).
Elliptical galaxies typically have intrinsically smooth surface
brightness profiles, unmarred by structural components such as spiral
arms or dust lanes.  However, their light is generated by a finite
number of stars, and the closer the galaxy is to us, the fewer of the
red giant stars that dominate the optical light will fall within each
seeing element of an image.  This number of stars will be subject to
Poisson statistics, and therefore the closer the galaxy is to us, the
rougher the image will appear.  This roughness is therefore a
measurement of the distance of the galaxy.  This is a method that
holds a tremendous amount of promise for peculiar velocity work, as it
yields distances to accuracies of 5\% for at least some
galaxies.  Unfortunately, the method is seeing limited, and requires the
resolution of the Hubble Space Telescope to reach much beyond 3000
\kms.  Although there now exist substantial data for nearby galaxies
using this technique (e.g., Tonry \etal\ 1997), it has not yet been
exploited for studies of the peculiar velocity field in any detail. 

\b The most luminous galaxies in clusters have long been recognized
to be distinct from the ``ordinary'' ellipticals that
make up the bulk of the galaxy population in ellipticals.  Gunn \& Oke
(1975) and Postman \& Lauer (1995) show that the luminosity of Brightest
Cluster Galaxies within a metric aperture is directly related to the
logarithmic slope of their surface brightness profile; using this
yields a distance to an accuracy comparable to that the TF relation,
15-20\%.  These have been used to measure the bulk component of the
velocity field on very large scales (Lauer \& Postman 1994); more on
this below. 

\b Finally, it has long been recognized that Type 1a supernovae are
standard candles (e.g., Phillips 1993).  When one includes information
about the shape of the light
curve, the peak luminosity is predicted well enough to yield distances
accurate to 5\% (Hamuy \etal\ 1996; Riess, Press \&
Kirshner 1996).  Supernovae are of course rare events, and therefore
one cannot {\it a priori} pick one's sample of galaxies 
via this technique; nevertheless, there have been a
number of recent measurements of the peculiar velocity field from
extant supernovae data (e.g., Riess, Press, \& Kirshner 1995, Zehavi
\etal\ 1998). 
\end{itemize}

\begin{figure*}
 \centering\mbox{\psfig{figure=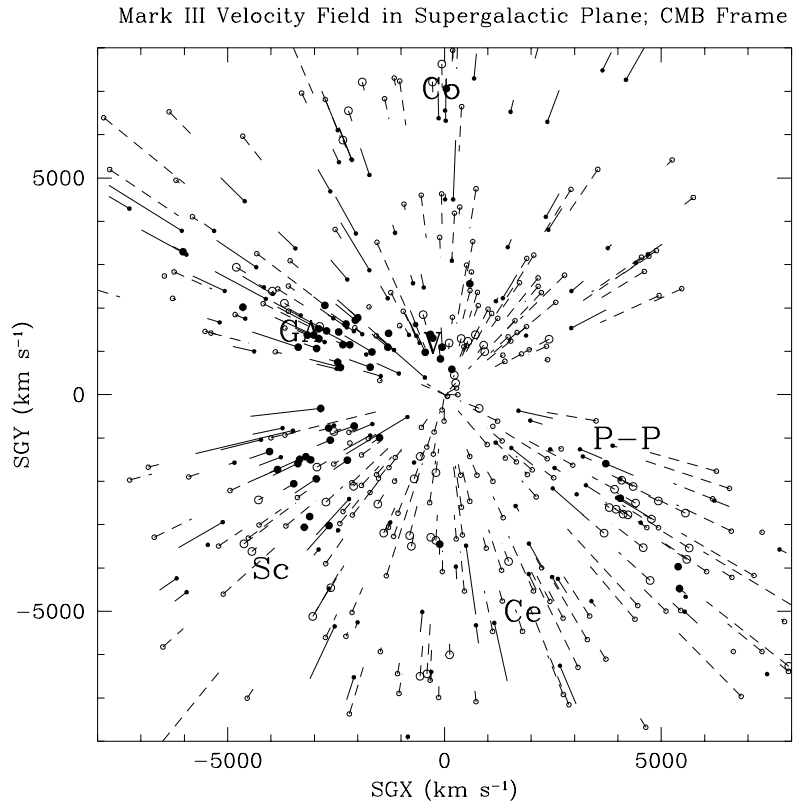,height=11cm}}
\caption[]{The Mark III peculiar velocities of all galaxies within
22.5$^\circ$ of the Supergalactic plane. The point is drawn at the
measured distance of the galaxy, while the line is drawn to its
redshift, in the CMB rest frame.  Positive peculiar velocities are
drawn with solid points and solid lines, while negative peculiar
velocities use open points and dashed lines. Points representing
groups of more than three galaxies are drawn somewhat larger.  Some of
the large superclusters in the nearby universe are marked.}
\label{fig:MarkIII}
\end{figure*}

Although the TF and fundamental plane relations are less accurate than
surface brightness fluctuations or Supernovae 1a's, they 
have been the distance indicators most widely used for
mapping the nearby velocity field, mostly because they are applicable
to the largest numbers of galaxies, and therefore can be used to
describe the velocity field in detail.  A large number of surveys of
nearby galaxies have used these techniques (SW; Postman
1995; Strauss
1997; Willick 1998).  Willick \etal\ (1995, 1996, 1997a) have combined
the data from a number of these, and put them on a common footing, to
create the Mark III catalog of peculiar velocities,
which covers much of the sky to a depth of roughly 6000 \kms.
Haynes \etal\ (1998) have carried out a TF survey of spiral
galaxies with somewhat poorer sampling, but with fewer difficulties 
concerning
matching disparate data sets, to create their SFI data set.  We 
discuss the results of these two surveys in some
detail below.  Figure~\ref{fig:MarkIII} shows the peculiar velocity
field of the Mark III dataset, as projected onto the Supergalactic
Plane.  Dots are drawn at the measured distance of the galaxy, and the
line ends at its measured redshift (in the CMB frame); thus the length
of the line is the measured radial peculiar velocity.  To reduce
errors, and the clutter in the diagram, neighboring galaxies have been
grouped following an algorithm described in Willick \etal\ (1996).

  There is clearly a tremendous amount of information in a plot like
this, but the data shown are noisy and sparse, and careful thought is
needed to use it in a statistically rigorous way.  One statistic that
one would like to measure is the coherence scale of the velocity
field.  That is, one can measure the mean bulk flow within a sphere
centered on the Local Group; the Cosmological Principle implies that
within large enough a sphere, this flow should be zero.  The
question, of course, is on what scale this is the case.  We will not
review this rather controversial subject here, other than to point out
that the situation right now is particularly confused (and therefore
exciting!): the Lauer-Postman (1994) measurement of bulk flow of order 700 \kms\ on
scales of 15,000 \kms\ remains unconfirmed, although despite much
scrutiny, nobody has found any fault at all with the data or the
analysis.  Meanwhile, the SMAC survey is finding a bulk flow of a
similar amplitude on similar scales, although in a quite different
direction from Lauer-Postman (Smith and Hudson, these proceedings).
On the other hand, Giovanelli \etal\ (1998a,b) are finding impressive
evidence from their field and cluster TF survey for negligible bulk
flows within a sphere of radius 5000 \kms\ centered on the Local
Group.  As Strauss (1997) emphasizes, a clean answer to this question
will have much to say about the mass power spectrum on large scales.

In this review, we would like to focus on a different subject, the
comparison 
of the peculiar velocity data with the galaxy density field via
equation~(\ref{eq:v-integral}), and its implications for bias and $\Omega$.

\section{The Relative Distribution of Galaxies and Dark Matter}

Equation~(\ref{eq:v-integral}), or its differential form,
equation~(\ref{eq:del-dot-v}), make reference to the {\em mass}
density fluctuation field $\delta$.  We would like to compare the
observed peculiar velocity field with the observed galaxy density
field, as determined, e.g., from redshift surveys of \iras\ galaxies
(Fisher \etal\ 1995; Yahil \etal\ 1991; Branchini, these proceedings).
This requires a model for 
the relation between the galaxy and dark matter distributions.  

The density fields $\rho$ of galaxies and dark matter of course differ by their relative
contributions to the overall density of the universe, so we work with
the density fluctuation field $\delta$ (equation~\ref{eq:delta}),
smoothed on some scale $R$. 

The simplest model, which was implicitly assumed in the early
days of the subject of large-scale structure, was that the two density
fluctuation fields were identical: 
\begin{equation} 
\delta_{\rm galaxies} = \delta_{\rm dark\ matter}. 
\label{eq:nobias}
\end{equation}

In 1984, Kaiser wrote a tremendously important paper, pointing out
that if clusters of galaxies form preferentially in denser regions
of space, their clustering would be stronger, or {\em biased},
relative to the galaxies.  It was quickly realized that the same
statement could hold for galaxies relative to dark matter.  Although
Kaiser's original formulation really only made statements about the
strength of the two-point correlation function, it was quickly made
more specific with the so-called {\em linear bias model}:
\begin{equation} 
\delta_{\rm galaxies} = b\,\delta_{\rm dark\ matter}.
\label{eq:linear-bias} 
\end{equation}
This became a quite important idea, as it gave modelers a crucial
extra degree of freedom with which to fit cosmological theories to the
increasingly stringent constraints from theory.  For example, Davis
\etal\ (1985) showed that the standard $\Omega = 1$ Cold Dark Matter
model could not be reconciled with existing data without invoking an
appreciable bias, $b \approx 2.5$.  Of course, this value of $b$ was
found to be in strong disagreement with that implied by the COBE
normalization, which is one of many ways of expressing the
non-viability of this model.  

  In addition to giving the theorist an extra parameter to play with,
Kaiser's bias idea made us realize that there was no reason 
{\it a priori} to expect
that  the distribution of galaxies and dark matter trace one
another perfectly.  The process of galaxy formation is complicated,
as the various detailed models presented at this conference by
Kauffmann, Lacey, and others indicate.  Therefore, the linear
proportionality in equation~(\ref{eq:linear-bias}), with a universal
bias parameter $b$ that is independent of position, scale, and time,
is almost certainly an oversimplification.  We could more generally
write down a generic relation between the two density fluctuation
fields: 
\begin{eqnarray} 
\nonumber \delta_{\rm galaxies} &=& {\cal F}(\delta_{\rm dark\ matter}) \\
&& {}+ \epsilon_{\rm dark\ matter},
\label{eq:general-bias} 
\end{eqnarray}
where $\cal F$ is a general, nonlinear function, which may depend on
smoothing scale $R$, and $\epsilon$ is a ``random''
variable. $\epsilon$ 
contains all the physics of galaxy formation which does not depend on
the local density $\delta_{\rm dark\ matter}$.
It turns out that even with
the freedom introduced by this general form, one can still say useful
things about the dark matter distribution (Dekel \& Lahav 1998), and
we will come back at the end of this review to a discussion of what
the properties of ${\cal F}$ and $\epsilon$ might be.  But for the
moment, let us stick with the simplifying assumption of linear
biasing, equation~(\ref{eq:linear-bias}).  In this case, comparison of
the peculiar velocity field with the {\em galaxy} density fluctuation
field via equation~(\ref{eq:del-dot-v}) yields: 
\begin{equation} 
\nabla \cdot \bfv(\bfr) = - {\Omega^{0.6} \over b} \delta_{\rm galaxies}(\bfr),
\label{eq:del-dot-v-b} 
\end{equation}
and similarly for equation~(\ref{eq:v-integral}).  Thus direct
comparison of peculiar velocity and galaxy density fields allows us to
constrain not $\Omega$ directly, but rather the degenerate combination
 $\beta \equiv {\Omega^{0.6}/ b}$. 

\section{Comparing Peculiar Velocities with the Galaxy Density Field} 

As indicated above, the noisiness of existing peculiar velocity data
makes analyses using equation~(\ref{eq:del-dot-v-b}), or its integral
counterpart, quite difficult, requiring much care to control subtle
systematic effects.  The whole field is reviewed in SW; we here
describe some of the recent approaches to the problem, emphasizing
where their results diverge. 

\subsection{Density-Density Comparison}

We would like to use equation~(\ref{eq:del-dot-v-b}) to make the
comparison between the velocity and density fields.  Unfortunately,
our peculiar velocity data give only a noisy realization of just the
radial component of the velocity field at a sparse and inhomogeneous
set of points.  Dekel \etal\ (1990) and Bertschinger \etal\ (1990)
have developed a method, called POTENT (subsequently refined by Dekel
1994; Dekel \etal\ 1998) to get around this problem.  The fundamental
insight comes from the assumption (well-justified on large scales)
that the velocity field exhibits potential flow, so that it may be
expressed as the gradient of a potential $\Phi$:
\begin{equation} 
\bfv(\bfr) = - \nabla \Phi(\bfr).
\label{eq:potential} 
\end{equation}
In this case, the radial component of the velocity field alone
contains enough information to determine $\Phi$.  In particular, 
integrating the observed radial component of the velocity field along
radial rays yields $\Phi$, and the gradient of the resulting
field yields the full three-dimensional velocity field.  A further
divergence yields the quantity on the left-hand-side of
equation~(\ref{eq:del-dot-v-b}), which can be compared directly with
the observed $\delta_{\rm galaxies}$ from a redshift survey. 

This program has been carried out recently by two groups.  Hudson
\etal\ (1995) used an early version of the quantity $\nabla \cdot
\bfv$ derived from the Mark III data to compare with Hudson's (1993)
reconstruction of the density field of optically selected galaxies.
They found good qualitative agreement between the two fields, and
derived a value $\beta_{optical} = 0.74 \pm 0.13$, where the
$\scriptstyle optical$ subscript refers to the bias value for
optically selected galaxies.  However, they could not fully explain
the observed scatter in their $\nabla \cdot \bfv -\delta$ comparison, given
the known errors in each of those quantities separately, and were
forced to invoke additional sources of scatter. 

\begin{figure*}
 \centering\mbox{\psfig{figure=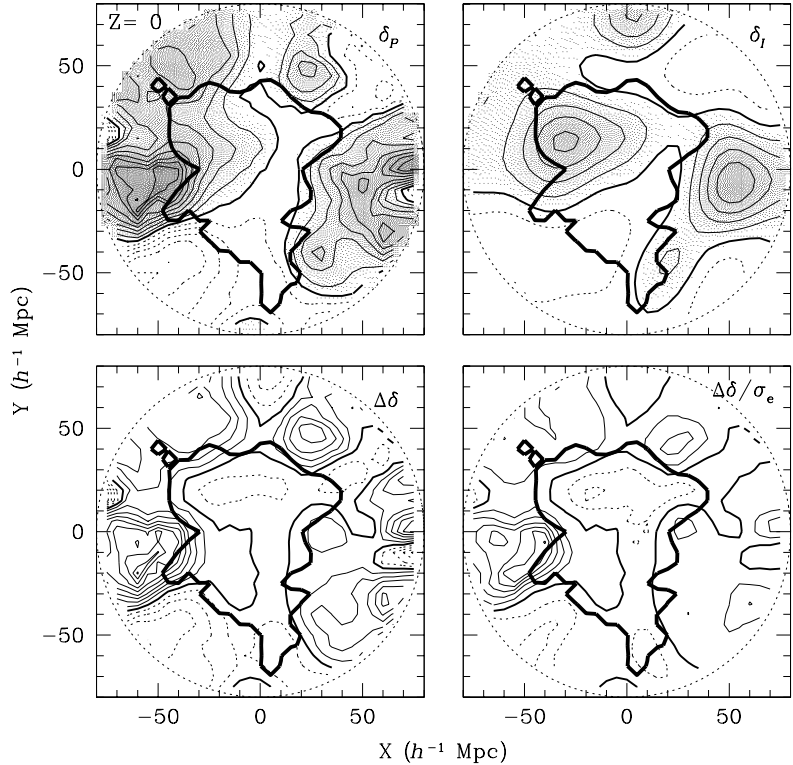,height=11cm}}
\caption[]{The density fluctuation field in the Supergalactic Plane at
1200 \kms\ smoothing, as determined from the POTENT analysis (upper
left-hand corner) and the \iras\ 1.2 Jy redshift survey (upper
right). Contours are separated by 0.2 in $\delta$. 
The lower left panel shows the difference between the two fields
(after linear scaling). The lower right shows the difference,
divided by the estimated errors; contours are separated by unity.  The
heavy contour marks the region within which the POTENT errors are
well-understood.}
\label{fig:potiras}
\end{figure*}

  More recently, Sigad \etal\ (1998) used the latest version of the
POTENT density field to compare with the density field of \iras\
galaxies.  The comparison of the two density fields in the
Supergalactic Plane is shown in Figure~\ref{fig:potiras}.  The
effective smoothing of the two fields is a 1200 \kms\ Gaussian.  As
the peculiar velocity data get sparser and noisier, the systematic
errors in $\nabla \cdot \bfv$ get more difficult to control; the heavy
solid line that outlines a shape roughly like the subcontinent of
India represents that region within which these errors are
well-understood and finite, as determined from extensive Monte-Carlo
experiments.  The qualitative agreement between the two maps is
impressive.  We do a linear scaling of one relative to the other (see
below), take the difference, and divide by the estimated errors; the
result is the lower right-hand panel.  The most significant
discrepancy is found in the Zone of Avoidance (corresponding
approximately to $SGY = 0$ in this projection), near the Great
Attractor, at $SGX \approx -40 \mpc, SGY \approx 0$.  The peculiar
velocity data are of course very sparse there, and the POTENT errors
are therefore correspondingly high (causing the pinching in the waist
of the region within which the POTENT data are to be most trusted).
Nevertheless, the discrepancy here remains a serious area of concern,
and requires further work.

  If we restrict ourselves to the region of space within which we
claim to understand our errors (and in particular, stay away from the
Kashmir\footnote[3]{Thanks to Luiz da Costa for this terminology!} of the
discrepant region near the Great Attractor), and do a linear
regression of $\nabla 
\cdot \bfv$\footnote[4]{Actually, the analysis uses a nonlinear extension
of this linear-theory expression; see Sigad \etal\ (1998) for
details.} on $\delta_{IRAS}$ on a grid with 500
\kms\ spacing (taking into account the errors in both dimensions), we
find a slope of $0.89$.  More importantly, we can define a statistic $S$
like a $\chi^2$/dof, which relates the residuals from the fit to the
estimated errors per point; we find a value $S = 1.06$.  We have
carried out Monte-Carlo experiments which include 
all relevant selection effects and sources of error to check the
expectation values of these quantities.  Our observed value of $S$
falls right in the range of those found by Monte-Carlo experiments,
thereby allowing us to conclude that
\begin{itemize} 
\b we understand our errors, and 
\b the data are consistent with the null hypotheses of
gravitational instability theory and linear biasing. 
\end{itemize}
The Monte-Carlo experiments were based on an $N$-body
simulation with an effective $\beta = 1$; we found that the method
delivers an unbiased result, with an error of order 0.12, so we
conclude further that
\begin{equation} 
\beta_{IRAS} = 0.89 \pm 0.12.
\label{eq:potent-answer} 
\end{equation}
We are in the process of carrying out similar experiments with
both $\Omega = 1$ and $\Omega = 0.3$ simulations with $\beta = 0.3$,
and we find that we get unbiased results for $\beta$ from our analysis
in this case as well, so we can say with some confidence that the data are
{\em inconsistent} with $\beta = 0.3$. 

\subsection{Velocity-Velocity Comparison}

Another approach to the determination of $\beta$ involves the integral
form of the velocity-density relation, equation~(\ref{eq:v-integral}).
In particular, this equation allows one to make a model for the
expected velocity field, given the observed galaxy density field from
a redshift survey; this may then be compared to observed peculiar
velocities.  In practice, one can avoid the various nasty Malmquist
biases that would otherwise come up in this game by phrasing the
problem slightly differently: establishing the TF relation requires
knowledge of the distances of galaxies to determine absolute
magnitudes.  Given measured redshifts, this requires a model for the
velocity field.  The correct model for the velocity field is that
which minimizes scatter in the TF relation (or, in somewhat more
sophisticated approaches, maximizes the likelihood of the TF
observables).  One important advantage of this approach is that it
works directly with the observables themselves, and therefore avoids
the great deal of data massaging that the POTENT analysis requires. 

This basic approach has been carried out by a number of
workers.  Shaya, Peebles,
\& Tully (1995) used an action principle extension of linear
theory to generate a velocity field model from Tully's (1987) catalog
of nearby galaxies, weighted by the blue light of each
galaxy.  Minimizing the scatter in the inverse form of the TF relation
for a somewhat massaged form of the Aaronson \etal\ (1982) data,
yielded a
value $\beta_{optical} =0.34 \pm 0.13$\footnote[5]{They expressed their
results in terms of $\Omega$, and explicitly assumed $b = 1$ for their
sample.}.  Their effective smoothing
length was quite small, only a few Mpc.

Davis, Nusser, \& Willick (1996) did the clever trick of expressing
both the observed (from Mark III) and modeled (from \iras) velocity
fields in terms of the same spherical-harmonic-based orthonormal
expansion.  They then developed a formalism that allowed them to
express the inverse TF relation in terms of this expansion.  This has
the great advantage of smoothing the data on large scales to ensure
linearity, and guarantees that the two fields are smoothed on
equivalent scales.  To their surprise, they found that within a
distance of 5000 or 6000 \kms, the two velocity fields disagreed in
even the lower-order multipoles.  They suggested that there are
possible systematic errors in the matching of the different peculiar
velocity datasets which make up the Mark III catalog, and therefore
they do not quote a value of $\beta$.

da Costa \etal\ (1998) used the same analysis as Davis \etal\ (1996), 
now using the SFI TF sample of Haynes \etal\ (1998), which should
be less subject to the possible matching problems of the Mark III.
They found very good qualitative agreement between the velocity
fields, and concluded that $\beta_{IRAS} = 0.6 \pm 0.1$.  The final TF
scatter was unfortunately appreciably larger than expected {\it a
priori}, perhaps due to components of the velocity field on scales
smaller than they were modeling. 

Finally, Willick \etal\ (1997b) and Willick \& Strauss (1998) carried
out a rigorous likelihood analysis of the Mark III TF data (termed
VELMOD). Unlike the
other approaches, it allows one to explicitly take into account the
presence of regions in which the redshift-distance relation along a
given line of sight is non-monotonic (triple-valued regions), the
scatter due to small-scale unmodeled components of the velocity field,
variations of the TF scatter with luminosity, variations in the TF
calibration from one dataset to another, and other systematic effects.
VELMOD requires a {\em full} modeling of the velocity field, and
thus the effective smoothing of the analysis was pushed to as small
scales as possible; 400 \kms\ was used in practice.  The Mark III and
\iras\ data were used here; they were found to be consistent with the
linear biasing and gravitational instability model.  The formal
analysis gives a value $\beta_{IRAS} = 0.50 \pm 0.04$, where the
impressively small statistical error bar is confirmed by Monte-Carlo
experiments.  Interestingly, this value is in excellent agreement with
the value above for optical galaxies by Shaya \etal\ (1995), when one
takes into account the relative bias of optical and \iras\ galaxies of
1.4 (e.g., Hermit \etal\ 1996).   The VELMOD analysis confirms the
suspicion of Davis \etal\ (1996) of systematic errors in the matching
of the datasets that make up Mark III; the best-fit TF relations
differ systematically from those assumed in Davis \etal\ (1996),
in such a way to explain the discrepancies the latter found in their
analysis.  With the freedom to fit the TF relation, Willick \& Strauss
(1998) found excellent
agreement between the Mark III and \iras\ datasets. 

So the velocity-velocity comparisons carried out here are in
substantial agreement with one another.  The density-density
comparison using the POTENT map gives a value of $\beta_{IRAS}$, {\em
using the same data}, that differs by several sigma.  Those who would
argue for $\Omega = 1$ will be buoyed by one set of analyses, while
those who would argue for $\Omega = 0.3$ will be cheered by the other.
Why the discrepancy?  Several possibilities come to mind:
\begin{itemize} 
\b The systematic sampling errors in the Mark III propagate into
the POTENT analysis to bias the determination of $\beta_{IRAS}$.
However, preliminary tests show that the POTENT density field is
actually quite robust to these problems.  The POTENT density field is
a local quantity; it should be insensitive to these global problems in
the velocity field. 
\b Both the POTENT and VELMOD codes have been extensively tested
with Monte-Carlo experiments based on PM simulations of the local
universe, following Kolatt \etal\ (1996).  It is possible that the
finite force resolution of the code is underestimating non-linear
effects on small scales, which may have unknown effects especially on
the VELMOD analysis, with its appreciably smaller smoothing scale.  We
are planning simulations at higher resolution to test this
possibility.  Willick \& Strauss (1998) show that their results are in
fact quite robust to a substantial increase in smoothing scale. 
\b Again, VELMOD and POTENT use quite different smoothing scales.
If the effective bias is a function of scale, one expects the
value of $\beta$ in the two analyses to differ.  Moreover, stochasticity in the
galaxy-mass relation (i.e., the $\epsilon$ term in
equation~\ref{eq:general-bias}) can cause {\em systematic} errors in
the determination of $\beta$ (Dekel \& Lahav 1998). Considering only
second moments of the galaxy-mass relation, the regression of
$\nabla \cdot \bfv$ on $\delta_{\rm galaxies}$ has a slope of
$r\Omega^{0.6}/b$, where $b\equiv\sigma_g/\sigma$ is the ratio of
the variances in the galaxy and mass density fields and 
and $r\equiv\langle\delta_g\delta\rangle/\sigma_g\sigma$ is the correlation
coefficient.  Note that VELMOD, being a velocity-velocity comparison,
uses an integral of the density field over a large range of scales, so
it is more difficult to understand the effect of stochasticity and
scale-dependent bias on these results.  We plan to use realistic bias
models (see below) to create simulations of their effect on VELMOD. 
\end{itemize}

\subsection{Other Approaches to $\beta$ and $\Omega$ from Peculiar Velocities}

There are a number of
other ways to constrain $\beta$ and $\Omega$ from peculiar velocity
data and equation~(\ref{eq:del-dot-v}).  
The redshifts of galaxies differ from the distances by
the radial component of the peculiar velocity,
equation~(\ref{eq:Hubble2}).  Because only the radial component of the
inferred three-dimensional position of a galaxy in redshift space is
affected, the clustering of galaxies acquires a systematic radial
anisotropy. Consider a cluster of galaxies.  On small scales, the
virial motions within the cluster cause what is a compact structure 
in real space to appear stretched out along the line of
sight in redshift space, thus {\em reducing} the apparent strength of
the clustering.  On large scales, coherent infall of galaxies towards
this overdensity makes galaxies appear closer to the cluster in
redshift space than in real space, {\em enhancing} the clustering
strength.  Hamilton (1998) has written a comprehensive review of
attempts to measure this large-scale effect from redshift surveys; its
amplitude can be related to the strength of clustering using linear
perturbation theory, yielding an estimate of $\beta$.  His summary of
published analyses is $\beta_{IRAS} = 0.77 \pm 0.22$, but individual
measurements are noisy, with large errors.  More importantly,
measurements of the anisotropy of the correlation function or power
spectrum show systematically lower values of $\beta$ than do analyses
measuring redshift-space distortions in spherical harmonics,
reminiscent of the density-density vs.~velocity-velocity dichotomy in
results described above. 

  Dekel \& Rees (1994) realized that one can put a firm lower limit on
$\Omega$, independent of $\beta$, from peculiar velocity data alone.
By definition, all regions of space have $\delta \ge -1$.  In linear
theory, we can write $-\nabla \cdot \bfv = \Omega^{0.6} \delta \ge
-\Omega^{0.6}$.  Therefore, the {\em lowest} observed value of $-\nabla
\cdot \bfv$ allows us to put a lower limit on $\Omega$.  Using the
Sculptor Void ($SGX = -20\mpc, SGY = -40\mpc$) in the Mark III POTENT maps
yields a limit $\Omega > 0.3$ at the 2.4$\,\sigma$ confidence level.
This result needs further checking with simulations \`a la Kolatt
\etal\ (1996), and with the latest versions of the Mark III and SFI
data.  

  One can use the POTENT maps in another way to put constraints on
$\Omega$.  In inflationary models, one expects the initial one-point
density distribution function to be Gaussian, while non-linear
gravitational growth causes systematic deviations from Gaussianity.
Nusser \& Dekel (1993) have developed a ``time machine'' with which they
can correct the quantity $\nabla \cdot \bfv$ for non-linearities, to
regenerate the original linear density field, and therefore its
distribution function.  This time machine depends on $\Omega$; they
find that with $\Omega = 1$, the initial distribution function is
beautifully Gaussian, while it is far from Gaussian for $\Omega =
0.3$.  They put a limit $\Omega > 0.3$ at $4\,\sigma$ confidence with
an early version of the Mark III data, under the assumption of
Gaussian initial conditions.  Again, this needs checking with modern
data, with special attention paid to systematics, which may effect the
tails of the distribution. 

Finally, one can try to measure the power spectrum of density
fluctuations that gives rise to the peculiar velocity field that one
sees.  Expressing equation~(\ref{eq:del-dot-v}) in Fourier space,
\begin{equation} 
i \bfk\cdot \tilde \bfv(\bfk) = \Omega^{0.6} \tilde \delta(\bfk),
\label{eq:v-delta-k} 
\end{equation}
one sees that the velocity field is more sensitive to large-scale waves
(i.e., smaller $k$) than is the density field.  Therefore, the
velocity field is a particularly effective way of measuring the power
spectrum on large scales (cf., the discussion in Strauss 1997).
Kolatt \& Dekel (1997) have measured the power spectrum of $\nabla
\cdot \bfv$ directly, while Zaroubi \etal\ (1997) and Freudling \etal\
(1998; see also Zehavi, these proceedings) have used a likelihood
technique to fit the ``raw'' peculiar velocity data for the power
spectrum.  Of course, the quantity they end up constraining is not
$P(k)$ directly, but rather $\Omega^{1.2} P(k)$.  Comparing with
models, and incorporating constraints from COBE and redshift surveys,
then leads to a constraint on $\sigma_8\Omega^{0.6}$, or equivalently
$\beta$.  These various papers are substantially in agreement; the
latter paper yields $\beta_{IRAS} = 1.2 \pm 0.2$. 

\section{The Complications of Bias}

As we've hinted at above, one possible explanation of these 
disparate results is that our simplistic model of linear,
deterministic, scale-independent bias is not valid.  
The analyses above differ substantially in their effective smoothing
scales as well as in the morphology of the galaxies which they observe. In
addition, the methods they use depend differently on the degree of
stochasticity in the galaxy-mass relationship (Dekel \& Lahav
1998). While these issues are unimportant if linear bias holds, a
nonlinear, stochastic, morphology-dependent bias could easily cause
the results of these analyses to differ.
To understand how a realistic bias affects matters, 
Blanton \etal\ (1998) 
have examined the relative distribution of galaxies and dark matter in
a simulation which models both the gravitational physics of dark
matter and the gas physics of the baryons. The simulation handles star
formation by converting gas into collisionless particles in regions
with infalling gas, with cooling times below the local dynamical time,
and masses above the Jeans mass. One can then look at the relationship
between the density field of these collisionless particles (which we
take as a proxy for the galaxy density field) and the dark matter
density field.

\begin{figure*}
\centering\mbox{\psfig{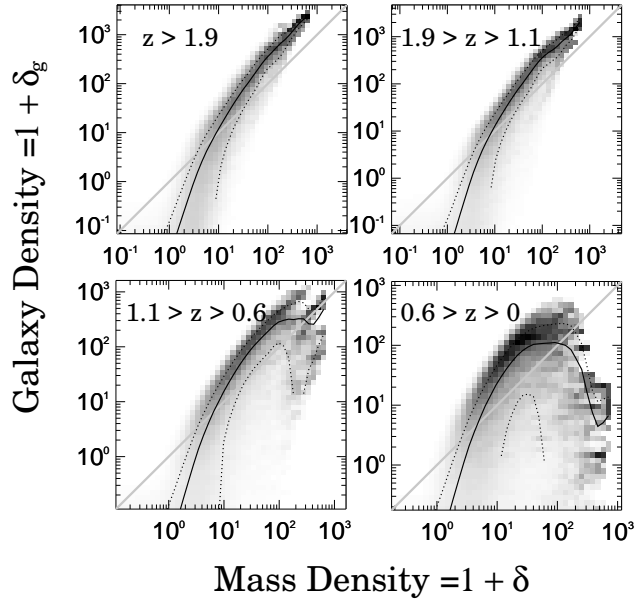}}
\caption{Galaxy mass density as a function of dark matter density for
each age quartile, at 1 $h^{-1}$ Mpc radius top hat smoothing.  Each
panel lists the range of formation redshifts included. Shading is a
logarithmic stretch of the conditional probability $P(1+\delta_g |
1+\delta)$.  Solid lines indicate $\vev{1+\delta_g|1+\delta}$; dotted
lines indicate the 1$\sigma$ deviation from the mean.}
\label{fig:delta-delta}
\end{figure*}

The resulting bias relationship is nonlinear,
stochastic, and is a strong function of galaxy {\it age}.  These
properties are revealed in Figure~\ref{fig:delta-delta}, which shows as a greyscale the
conditional probability $P(1+\delta_g|1+\delta)$ and as the solid line
the conditional mean $\vev{1+\delta_g|1+\delta}$, where all quantities are
defined with a top hat filter of radius 1 $h^{-1}$ Mpc.  Each panel
shows the results at $z=0$ for galaxies formed at different epochs, as
labeled.  Note that the oldest galaxies are the cleanest tracers of
the dark matter distribution: the scatter around the mean
galaxy-mass relation is small. However, the youngest
galaxies show a nonlinear, even non-monotonic, relation with the
dark matter; they are under-represented in the very densest regions of
the dark matter map (reminiscent of spirals in the cores of
clusters, although in the real universe clusters still represent
appreciable overdensities in the distribution of late-type galaxies;
Strauss \etal\ 1992). Also, the scatter around the mean
density relation for the youngest galaxies is large.

\begin{figure*}
\centering\mbox{\psfig{figure=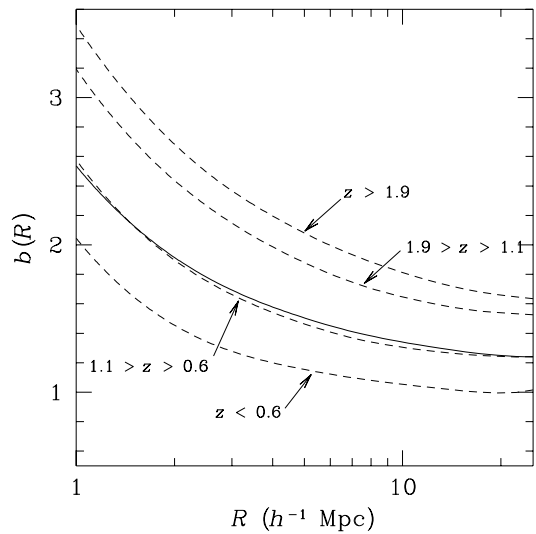,height=8cm}}
\caption{The bias $b(R)\equiv\sigma_g(R)/\sigma(R)$, where $R$ refers to
the top hat smoothing radius. Solid
line indicates all galaxies. Dashed lines indicate each age
quartile, with range of formation redshifts listed. Note the strong
scale-dependence, and that old galaxies are more biased than
young.}
\label{fig:bias}
\end{figure*}

In these simulations, the relationship between galaxies and mass also
depends on scale. In Figure~\ref{fig:bias}, we show the bias
$b\equiv\sigma_g/\sigma$ calculated on various scales.  The obvious
scale-dependence of $b$ is due to the dependence of the galaxy
formation process on temperature. The temperature sets the local Jeans
mass, which partly determines whether star-formation occurs: the
higher the temperature, the greater the overdensity needed to form
stars. On small scales the temperature is proportional to the
gravitational potential $\phi$. Note that in Fourier space, ${\tilde
\phi}(k) \propto {\tilde \delta}(k)/k^2$. For high $k$, then, there is
little power in the potential or temperature fields; {\it i.e.}~these
fields are {\it smoother} than the density field.  Thus, 
temperature correlates over large scales; furthermore, on these large
scales it correlates with density as well.
Thus the dependence of galaxy
formation on temperature can couple the galaxy density on small scales
with the dark matter density on larger scales. As 
Blanton \etal\ (1998) show, this coupling causes scale-dependence of
the bias  
relation. The dependence of galaxy formation on local gas temperature
is likely to be important in any galaxy formation scenario; thus, this
scale-dependence may be generic.

The work ahead is evaluate the consequences of this complicated bias
relationship for statistical measures of large-scale structure.  The
data definitely allow stochasticity; in the POTENT-\iras\ comparison
of Sigad \etal\ (1998), although the value of $\chi^2$/dof is
consistent with deterministic bias, it is also consistent with the
inclusion of the rms value of $\epsilon$ found in the simulations at
1200 \kms\ smoothing.  Moreover, nonlinear relative bias between
galaxies of different types is unambiguously observed, in the form of
the morphology-density relation (cf.\ the discussion in Strauss \&
Blanton 1998). 
We are in the process of carrying out simulations using realistic bias
laws, to determine how it may skew the results of the various
analyses described above.

\section*{Acknowledgments}

We acknowledge our colleagues on the various work described here: the
POTENT analysis (Yair Sigad, Ami Eldar, Avishai Dekel, and Amos
Yahil), the VELMOD analysis (Jeff Willick, Dekel, and Tsafrir Kolatt),
and the measurement of bias from simulations (Renyue Cen and Jerry
Ostriker).  This work was supported in part by the Alfred P.~Sloan
Foundation, Research Corporation, NSF grant AST96-16901, and the Princeton
University Research Board.

\end{document}